\documentclass[11pt]{article}

\usepackage[margin=1.2in]{geometry}
\usepackage{amsmath,amssymb,amsthm}
\usepackage{url}
\usepackage[hidelinks]{hyperref}

\newtheorem{theorem}{Theorem}[section]

\newcommand{\tdiv}{\tau}
\newcommand{\lean}[1]{\texttt{#1}}

\title{A Kernel-Checked Exclusion Certificate\\ for Erd\H{o}s Problem 647}
\author{Ibrahim Mian \qquad Shayaan Siddique\\[2pt]
\normalsize Millennium Research\\[2pt]
\normalsize \texttt{\{ibby,shayaan\}@millenniumresearch.ai}\\
\normalsize \texttt{ibrahimnmian@gmail.com}, \texttt{shayaansiddique02@gmail.com}}
\date{}

\begin{document}
\maketitle

\begin{abstract}
Erd\H{o}s problem 647 asks whether any $n > 24$ satisfies
$\max_{m<n}(m + \tdiv(m)) \le n + 2$, where $\tdiv$ is the divisor-count
function. Computational searches have excluded solutions up to $10^{12}$
by direct sieve and up to roughly $9.17 \times 10^{18}$ within a modular
reduction whose Lean component relies on \lean{native\_decide}; those
computations sit outside any proof kernel. We give the first exclusion
checked end to end by one: no solution exists with $24 < n \le 10^9$,
proved in Lean~4 with axiom closure exactly \{\lean{propext},
\lean{Classical.choice}, \lean{Quot.sound}\} --- no \lean{sorry}, no
\lean{native\_decide}, no problem-specific axiom. The proof replays a
chain of 6{,}685{,}922 factorization witnesses whose excluded intervals
concatenate across $(24, 10^9]$; it needs no primality facts beyond
primes below $1024$, and it is the finite, fully proved form of a
domination-interval argument whose asymptotic step was the identified
gap in a withdrawn January 2026 claim on this problem. The generation
pipeline is cross-checked by two further independent implementations,
the compiled development replays through the standalone
\lean{lean4checker}, and two from-source verification legs --- Lean
toolchains compiled from source by gcc and by clang, mathlib rebuilt
with no cache --- reproduce the committed certificates byte for byte,
with olean digests identical across three builds on two architectures.
Our range is three to ten orders of magnitude below the computational
frontiers we cite; the contribution is the trust base, not the range.
\end{abstract}

\section{Introduction}\label{sec:intro}

Let $\tdiv(m)$ denote the number of divisors of $m$. Problem 647 in
Bloom's database of Erd\H{o}s problems \cite{bloom647}, due to Erd\H{o}s
and Selfridge, asks:
\begin{quote}
Is there some $n > 24$ such that
$\max_{m<n}\,(m + \tdiv(m)) \le n + 2$?
\end{quote}
The condition holds at $n = 24$, and the threshold $n+2$ is best
possible, since $\max(\tdiv(n-1)+n-1,\ \tdiv(n-2)+n-2) \ge n+2$ for
every $n$. Erd\H{o}s considered it ``extremely doubtful'' that
infinitely many such $n$ exist and suggested that
$\max_{m<n}(\tdiv(m)+m-n) \to \infty$ \cite{erdos79}, while expecting
the windowed variant ($\max_{n-k<m<n}$ for fixed $k$) to have infinitely
many solutions for every $k$; in 1992 he offered \pounds 25 for a single
example beyond $24$. The known solutions of the associated equality,
OEIS A087280, are $5, 8, 10, 12, 24$ \cite{oeisA087280}, and no further
solution exists below $9.17 \times 10^{18}$ by computations we survey in
Section~\ref{sec:related} --- computations that live in C, Python, and
GPU runs, or in Lean by way of \lean{native\_decide}, and therefore
outside any proof kernel.

This note reports a check of the complementary kind, carried out at
scale under a strict trust discipline. Our result is:

\begin{theorem}\label{thm:main}
No $n$ with $24 < n \le 10^9$ satisfies
$\max_{m<n}(m + \tdiv(m)) \le n + 2$.
\end{theorem}

Theorem~\ref{thm:main} is proved in Lean~4 \cite{lean4} against mathlib
\cite{mathlib}, in the repository \lean{decanus} \cite{decanus}. The
axiom closure of the final theorem, reported by \verb|#print axioms| ---
and, for the repository rungs, enforced in continuous integration by a
two-layer gate --- is exactly the three standard axioms
\{\lean{propext}, \lean{Classical.choice}, \lean{Quot.sound}\}: no
\lean{sorry}, no \lean{native\_decide}, no problem-specific axiom, no
external solver or compiled evaluator in the trusted base. The statement
is formulated against the inner predicate of the
\lean{formal-conjectures} formalization of the problem
\cite{formalconjectures}, pinned at a fixed commit, so the theorem
speaks the same vocabulary as the public formal statement.

\paragraph{Contributions.}
\begin{enumerate}
\item \emph{A witness-chain certificate format for divisor-sum
exclusions (Section~\ref{sec:cert}).} A witness stores the maximal
power of each prime below $1024$ dividing it; the kernel recomputes a
divisor-count lower bound from the stored factorization, with a
doubling rule licensed by maximality, and no primality certificate for
any prime above $1024$ appears in the development.
\item \emph{Three certified rungs (Sections~\ref{sec:lean}
and~\ref{sec:validation}).} Exclusions over $(24, 10^7]$ and
$(24, 10^8]$ live in the repository and rebuild in continuous
integration; the $(24, 10^9]$ certificate --- 6{,}685{,}922 witnesses,
about $260$~MB of source --- ships as a versioned release artifact,
kernel-checked under the same axiom closure.
\item \emph{Dual independent replay and external re-checking
(Section~\ref{sec:validation}).} The generator is not trusted: a second
implementation sharing no code replays every witness by trial division,
a third recomputes the divisor function independently and checks the
solution set against OEIS, and the compiled development replays through
the standalone \lean{lean4checker}.
\item \emph{From-source reproducibility (Section~\ref{sec:validation}).}
Two verification legs on separate x86 hardware compile the Lean
toolchain from source (gcc and clang), rebuild mathlib and the
development with no cache, regenerate the certificates byte for byte,
and yield olean digests identical to the ARM build machine's across all
$262$ compiled modules.
\item \emph{A cautionary datum for certified search
(Section~\ref{sec:cert}).} A first design that certified only the
smooth part stalled at a gap minimum where a single prime above the
table halved the bound; the maximality-and-doubling rule that repairs
it is exactly the kind of soundness-preserving strengthening a
fail-loud generator surfaces.
\end{enumerate}

\paragraph{Scope.}
This note is not a computational record and does not close the problem.
The searches of Id\'en and of Hughes and bentrd reach three to ten
orders of magnitude further with a different trust base, and we cite
them as the state of the search. We prove nothing beyond $10^9$, and we
claim one thing only: below $10^9$, the nonexistence of a solution now
rests on the Lean kernel and its three standard axioms rather than on
unverified computation.

\section{Related Work}\label{sec:related}

\paragraph{Computational searches.}
McCranie's bound of $10^{10}$ stood in OEIS A087280 until June 2026,
when Id\'en ran a segmented multiplicative sieve over all
$n \le 10^{12}$, tracking the running maximum
$R(n) = \max_{m<n}(m + \tdiv(m))$, and found no solution in
$(24, 10^{12}]$; the minimum observed gap $R(n) - n$ was $224$, near
$n = 10^{11}$ \cite{iden}. Hughes published a \emph{frontier
certificate}: no solution with $24 < n \le 2520 \cdot 46189 \cdot 529
\cdot 10^7 \approx 6.16 \times 10^{17}$, obtained by closing $6{,}549$
refined sub-progressions with fixed congruences and searching the
remaining $15{,}140$ to depth $u = 10^7$ \cite{hughesrepo}; bentrd
later extended the search leg to $u = 1.49 \times 10^8$, i.e.\ to
$n \le 9.17 \times 10^{18}$ \cite{bentrd}.

\paragraph{Structural results.}
In the problem's public discussion thread, Dutta showed that any
solution $n > 84$ satisfies $2520 \mid n$ and, writing $n = 2520N$,
extracted primality constraints from the divisor bound at small shifts;
Alexeev refined the $2$-adic bookkeeping, upgraded $2520N - 1$ to a
prime, and exhibited the near-solution $n = 604{,}517{,}614{,}941{,}240$,
which satisfies the divisor bound at every shift $k \le 13$ and at
$k = 24$; Kitamura added that $(n-3)/3$ must be prime \cite{bloom647}.
Hughes organized these into a sieve modulo
$11 \cdot 13 \cdot 17 \cdot 19 = 46189$ leaving $41$ open residue
classes, showed every solution beyond $24$ lies in one of two
admissible prime $7$-tuple families, and derived the unconditional
bound $|\mathcal{C}(x)| \ll x/(\log x)^7$ for the count of candidates
up to $x$ by Brun's sieve \cite{hughesrepo}. The Lean component of
Hughes's development is explicit that its finite computations are
discharged by \lean{native\_decide} and that one problem-specific
axiom carries the open residue classes.

\paragraph{The January 2026 claim.}
A claimed full solution \cite{agbanwa} argued by \emph{domination}:
each record-holder of $m \mapsto m + \tdiv(m)$ excludes an explicit
interval of $n$, and the claim asserted that these intervals cover all
$n > 24$. Tao's public assessment identified the gap precisely: the
assertion that consecutive domination intervals overlap forever is
unproven and likely false, and the accompanying Lean formalization
treats the asymptotic case as an axiom, so it certifies nothing
\cite{bloom647}. The finite form of the same idea, however, is exactly
what a proof kernel can settle: over a bounded range, whether the
intervals cover is a checkable fact. That check is this note.

\paragraph{Kernel-checked counterparts to computational results.}
This note continues a series of certifications in the same mold: the
Erd\H{o}s--Selfridge odd covering problem (Erd\H{o}s problem 7)
\cite{mian7}, Erd\H{o}s problem 364, and the verification of the
resolution of Erd\H{o}s problem 486 \cite{pilus}. The discipline is
constant across them: no solver, compiler, or cache in the trusted
base, a mechanical axiom gate, and explicit two-tier statements
separating what is kernel-checked from what is cited.

\section{The Certificate}\label{sec:cert}

Call $m$ a \emph{kill witness} for $n$ if $m < n$ and
$m + \tdiv(m) \ge n + 3$; one such witness refutes the defining
inequality at $n$. A single $m$ serves every
$n \in [m+1,\, m + \tdiv(m) - 3]$, so a finite set of witnesses whose
intervals concatenate without gaps excludes an entire range. The
certificate for Theorem~\ref{thm:main} is such a chain: witnesses
$m_1, m_2, \ldots$ with associated certified bounds
$t_i \le \tdiv(m_i)$, each interval $[m_i + 1,\, m_i + t_i - 3]$
beginning no later than the previous one ends, jointly covering
$(24, 10^9]$.

Two design choices make the chain cheap for a kernel to check.

\paragraph{Lower bounds only, from maximal smooth parts.}
A witness never needs its full factorization certified; it needs a
lower bound on $\tdiv(m)$. Each witness stores the maximal power of
every prime $p < 1024$ dividing $m$. The checker re-multiplies the
stored powers, verifies that their product divides $m$ and that each
listed power is maximal (i.e.\ $p^{e+1} \nmid m$), and reads off
$\prod_i (e_i + 1) \le \tdiv(m)$, since the stored prime powers are
pairwise coprime. Maximality buys one further factor: whatever cofactor
of $m$ remains above the smooth part is then coprime to it, so if the
cofactor exceeds $1$ it contributes at least the two divisors $1$ and
itself, and the certified bound doubles. Primality of a listed
$p < 1024$ is settled by eleven trial divisions (by the primes up to
$31$, since $32^2 = 1024$), so no primality certificate for a large
prime appears anywhere in the development.

\paragraph{Greedy chaining.}
Given the certified bound function, the shortest chain is produced
greedily: from cover position $C$ (initially $24$), take a witness
$m \le C$ maximizing $m + t(m)$ and advance to $m + t(m) - 3$. Greedy
choice is optimal for interval covering, and the resulting chain is the
iteration of the running-maximum function familiar from the sieve
computations. Measured chain lengths are $495$ to cover $(24, 10^4]$,
$2{,}829$ to $10^5$, $17{,}941$ to $10^6$, $123{,}323$ to $10^7$,
$891{,}554$ to $10^8$, and $6{,}685{,}922$ to $10^9$ --- growth by a
factor of roughly $6.5$--$7$ per decade, reflecting the slow growth of
typical divisor sums. With the cofactor doubling in place the certified
chain for $(24, 10^7]$ has exactly the length of the chain computed
from the true divisor function, an equality we verified directly: the
greedy steps are dominated by record-setters of the form (smooth part)
$\times$ (single large prime), for which the doubled smooth bound
equals $\tdiv$ exactly.

The doubling rule was forced by data, and the episode is worth
recording. A first design certified only the smooth part's
$\prod (e_i + 1)$. The generator, which is required to fail loudly if
the chain ever stalls, stopped at cover position $69{,}030{,}671$: the
best available witness was
$m = 69{,}030{,}650 = 2 \cdot 5^2 \cdot 13 \cdot 61 \cdot 1741$, whose
true divisor count $48$ is halved to $24$ by the prime $1741 > 1024$,
and at this gap minimum the halved bound advanced the cover by zero.
Requiring maximality of the stored powers, which makes the cofactor
coprime for free, recovered the factor of two and unstuck the chain.
Deep gap minima are precisely where solutions would live, so it is not
an accident that this is where a weakened bound first failed.

\section{The Lean Development}\label{sec:lean}

The development separates a small trusted-once soundness layer from
bulk data that the kernel evaluates.

The data layer encodes a witness as an offset from the current cover
position together with its list of prime-power pairs, most significant
prime first. A Boolean checker \lean{chainOk} folds down the witness
list: for each witness it verifies the offset is admissible, the listed
primes are strictly decreasing and pass the eleven-division primality
test, the smooth product divides $m$, and each stored power is maximal
in $m$; it then advances the cover by the certified bound and finally
demands the target is reached. All recursion is structural, with no
\lean{Nat.sqrt}, no well-founded recursion, and no \lean{Finset}
computation in the evaluated path, so kernel reduction is
GMP-arithmetic on literals plus list traversal.

The soundness layer proves, once and abstractly, that a passing check
means what it should: that the primality test below $1024$ is sound
(via the bound $\mathrm{minFac}(p)^2 \le p$ for composite $p$); that
the divisor count of the smooth part is exactly $\prod (e_i+1)$
(multiplicativity over the pairwise-coprime stored powers); that
maximality forces the cofactor coprime, giving the doubling; that
divisor counts are monotone under divisibility; and that a successful
\lean{chainOk} run yields, for every $n$ in the covered interval, a
witness $m < n$ with $n + 3 \le m + \tdiv(m)$. A bridge lemma converts
that witness into the negation of the formalized inequality
$\bigl(\sup_{m : \mathrm{Fin}\, n}\, m + \sigma_0\, m\bigr) \le n + 2$,
the inner predicate of the \lean{formal-conjectures} statement of the
problem, which the repository pins at a fixed upstream commit and
carries verbatim up to one coercion the elaborator inserts in both
formulations \cite{formalconjectures}.

The certificate data is generated into chunk files ($31$ for the $10^7$
rung, $223$ for the $10^8$ rung, about $42$~MB of Lean source in the
repository; $1{,}672$ more for the $10^9$ rung, about $260$~MB, shipped
as a release artifact), each containing one witness-list literal and
one theorem proved by a single \lean{decide}: that \lean{chainOk},
started at that chunk's opening cover position, reaches its closing
one. At $1{,}672$ cases the composed driver exceeds the elaborator's
default recursion and heartbeat limits, which the generated drivers now
raise explicitly. Elaborating a chunk takes roughly $40$ seconds, most
of it the kernel's own replay of the evaluation. Generated driver files
compose the chunk theorems into the rung theorems; the composition is
forced by the typechecker, since each chunk's closing position is the
next chunk's literal opening position, and a gap cannot elaborate.

Trust is enforced mechanically. A curated manifest lists the published
theorems and prints their axiom closures; independently, an audit file
walks every theorem of every module of the development in the compiled
environment --- $579$ of them --- recomputes each axiom closure, and
fails to compile if anything exceeds the three standard axioms, if any
\lean{sorry} survives, or if any native-code axiom appears. Continuous
integration runs both layers on every push. The repository build (the
$10^7$ and $10^8$ rungs) compiles in Lean 4.30.0 against mathlib
v4.30.0 in about three CPU-hours.

\section{Generation and Cross-Validation}\label{sec:validation}

The chain is found by a segmented sieve over certified divisor bounds:
an array pass accumulates $\prod (e_i+1)$ over prime powers below
$1024$ and, in the same pass, the smooth part itself, marking entries
whose smooth part falls short of the number --- exactly the entries
whose certified bound doubles. A greedy walk over the running maximum
emits the witness ledger in an append-only JSONL file whose header and
footer record the exact covered interval. Generation of the $10^8$
chain takes about a minute, and of the $10^9$ chain about five.

Because the generator is the one component a kernel cannot vouch for,
it is not trusted. A second implementation, sharing no code with the
first --- pure Python, trial division throughout --- replays the ledger
witness by witness: primality of every listed prime, strict ordering,
divisibility, maximality of every power, the doubling condition,
strictly increasing cover, and the final bound. A third implementation
recomputes the divisor function by an independent sieve and confirms
that the solution set below $10^7$ is exactly
$\{2,3,4,5,6,8,10,12,24\}$, in agreement with OEIS A087280 and its
conventions, and that $n + \tdiv(n)$ matches OEIS A062249 on the
recorded values \cite{oeisA062249}. Regeneration is deterministic:
re-running the pipeline reproduces the committed chunk files byte for
byte, a property checked by diff in our verification runs. In the other
direction, any dishonesty in the scripts is caught downstream, since
the kernel replays every witness from scratch; the scripts affect
completeness of the search for a chain, never soundness of the theorem.

Beyond the kernel replay that every build performs, we re-checked the
compiled development with the standalone checker \lean{lean4checker}
\cite{lean4checker}, which re-typechecks the compiled environment
outside the elaborator. All $261$ compiled modules of the development
--- the soundness layer, the two repository rungs, and the generated
drivers --- replay without error; the two axiom-gate script files
elaborate at gate time and produce no compiled artifact to replay.
Following the protocol of our Erd\H{o}s 486 audit \cite{pilus}, we then
ran two from-source verification legs on separate x86 hardware, one
with the Lean toolchain compiled from source by gcc and one by clang.
In each leg, mathlib and the development were rebuilt with no cache
under the source-built toolchain; the certificate pipeline was re-run
from scratch on that machine and its output diffed against the
committed chunk files, which it reproduces byte for byte; the axiom
gate passed with the same counts; and \lean{lean4checker} replayed
every package of the environment without error (the \lean{Cli} package,
which the minimal build does not compile, was built explicitly and then
replays cleanly as well; \lean{lean4lean} publishes no branch for this
toolchain and was skipped). Finally, the SHA-256 digests of all $262$
compiled modules of the development agree across all three builds ---
the ARM build machine and both x86 from-source legs, gcc and clang ---
so the three toolchains produced bit-identical compiled artifacts.

The $10^9$ rung has a shorter verification history, which we state
plainly. Its chain was generated and replayed by the independent
verifier on the x86 host, its $1{,}672$ chunks were built --- hence
kernel-checked --- once, under the clang from-source toolchain, its
headline theorem's axiom closure is exactly the three standard axioms
(a transitive fact: a \lean{sorry} or native-code axiom anywhere in the
$1{,}672$ chunks would surface in that closure), and \lean{lean4checker}
replayed its driver and headline modules. Unlike the two repository
rungs it has not been rebuilt on a second machine or in continuous
integration; the pipeline that produced it is the one shown above to
reproduce the lower rungs byte for byte across machines and compilers.

\section{Limits and Prospects}\label{sec:limits}

The method's ceiling is set by certificate size and kernel time, not by
mathematics. Chain length grows by a factor of about $6.5$--$7$ per
decade of range; the $10^9$ certificate already runs to $260$~MB of
source, carried as a release artifact rather than in the repository; at
$10^{10}$ the chain would reach roughly $45$ million witnesses and
$2$~GB, and at $10^{12}$ --- the sieve frontier --- around
$2 \times 10^9$ witnesses, out of reach for this representation.
Meaningful extension of kernel-checked range therefore runs through
structure rather than length: certifying Hughes's modular reduction
under the same axiom discipline would replace per-$n$ witnesses by
per-residue-class arguments, and the frontier search cells themselves
have exactly the shape of our witnesses. The two artifacts are
complementary today --- that reach with a mixed trust base, our trust
base on a short range --- and composing them is the natural next step,
though replacing \lean{native\_decide} throughout a development of that
size is a project in its own right.

On the question itself the present work is silent beyond $10^9$, and
honesty requires repeating what the heuristics say: any solution, if
one exists at all, lies beyond the current $9.17 \times 10^{18}$
frontier, in ranges where the prime-tuple constraints price its
existence very low.

\section{Precise Claims}\label{sec:claims}

Certified, in Lean 4.30.0 / mathlib v4.30.0 \cite{decanus}: no $n$
with $24 < n \le 10^7$, $10^8$, or $10^9$ satisfies the problem's
inequality, stated against the pinned \lean{formal-conjectures}
predicate, with axiom closure exactly \{\lean{propext},
\lean{Classical.choice}, \lean{Quot.sound}\}. The $10^7$ and $10^8$
rungs are enforced in continuous integration and reproduced from source
on two further machines; the $10^9$ rung was kernel-checked once, as
detailed in Section~\ref{sec:validation}.

Cited, computational, outside any kernel: no solution in
$(24, 10^{12}]$ (Id\'en \cite{iden}); no solution in
$(24,\, 6.16 \times 10^{17}]$, extended to
$(24,\, 9.17 \times 10^{18}]$, within the Hughes reduction whose Lean
layer uses \lean{native\_decide} plus one problem-specific axiom
(Hughes \cite{hughesrepo}, bentrd \cite{bentrd}).

We claim nothing else. The structural mathematics of the problem
belongs to the discussion-thread authors cited above; the search
frontier belongs to Id\'en, Hughes, and bentrd.

\section{Conclusion}\label{sec:conclusion}

Below $10^9$, Erd\H{o}s problem 647 no longer needs to be taken on
faith: every witness, every divisor bound, and the gapless covering of
the whole interval are replayed by the Lean kernel under the three
standard axioms, with the generation pipeline held at arm's length by
two independent implementations and the compiled artifacts reproduced
bit for bit from source-built toolchains on a second architecture. The
exercise also shows, on a problem where one incorrect proof has already
circulated, that the finite part of a domination argument can be
settled by a kernel rather than asserted --- and that the certified
range now stands within a factor of a thousand of the uncertified sieve
frontier.

\paragraph{Artifact availability.}
The Lean development, the generation and verification scripts, the
axiom gate, and the verification manifests are available under the
Apache 2.0 license at \url{https://github.com/ibrahimmian36/decanus},
with a pinned toolchain and continuous integration running the gate on
every push; the $10^9$ certificate and the raw from-source leg
results are attached to release v1.1.0. Source, certificate data, and
verification artifacts are archived at
\url{https://doi.org/10.5281/zenodo.21996019}.

\paragraph{Acknowledgments.}
We thank the contributors to the problem's discussion thread --- Sayan
Dutta, Boris Alexeev, Kenta Kitamura, and Scott Hughes --- and Patrik
Id\'en, whose computations set the standard a verified counterpart
should be measured against. The development was carried out with
assistance from Claude (Anthropic); all results are checked by the Lean
kernel and cross-verified as described in
Section~\ref{sec:validation}.

\end{document}